\documentclass[12pt,preprint,fleqn]{aastex}
\usepackage{graphicx}
\slugcomment{Submitted to {\it The Astrophysical Journal Letters}}

\begin{document}

\title{Photon-photon Absorption of Very High Energy Gamma-Rays from
Microquasars: Application to LS 5039}

\author{Markus B\"ottcher\altaffilmark{1}
\&
Charles D. Dermer\altaffilmark{2}}

\altaffiltext{1}{Astrophysical Institute, Department of Physics 
and Astronomy, Ohio University, Athens, OH 45701, USA}
\altaffiltext{2}{E. O. Hulburt Center for Space Research, Code 7653
Naval Research Laboratory, Washington, D.C. 20375-5352}

\begin{abstract}
Very high energy (VHE) $\gamma$-rays have recently been detected from 
the Galactic black-hole candidate and microquasar LS~5039. A plausible
site for the production of these VHE $\gamma$-rays is the region close
to the base of the mildly
relativistic outflow. However, at distances 
comparable to the binary
separation, the intense photon field of the 
stellar companion leads to substantial $\gamma\gamma$ absorption of 
VHE $\gamma$-rays.
If the system is viewed at a substantial inclination 
($i \ne 0$), this absorption feature will be modulated on the orbital 
period of
the binary as a result of a phase-dependent stellar-radiation 
intensity and pair-production threshold. We apply our results to LS~5039 
and find that 
(1) $\gamma\gamma$ absorption effects will be substantial if the
photon production site is located at a distance from the central
compact object of the order of the binary separation ($\approx 2.5\times 
10^{12}$ cm) or less;
(2) the $\gamma\gamma$ absorption depth will be largest at a few
hundred~GeV, leading to a characteristic absorption trough; (3) the
$\gamma\gamma$ absorption feature will be strongly modulated on
the orbital period, characterized by a spectral
hardening accompanying periodic dips of the VHE $\gamma$-ray
flux; and (4) $\gamma$-rays can escape virtually 
unabsorbed, even from within $\approx 10^{12}$ cm, 
when the star is located behind the production site
as seen by the observer.
\end{abstract}


\section{\label{intro}Introduction}

Recent observations \citep{aharonian05} of $\gtrsim 
250$ GeV $\gamma$-rays with the High Energy Stereoscopic System (HESS)
from the X-ray binary jet source LS 5039 establish that microquasars
are a new class of $\gamma$-ray emitting sources. These results 
confirm the earlier tentative identification
of LS 5039 with the EGRET source 3EG~J1824-1514 \citep{paredes00}. 
In addition to LS~5039, 
the high-mass X-ray binary LSI~$61^o303$  (V615~Cas) 
also has a possible $\gamma$-ray counterpart in the MeV -- GeV
energy range \citep{gregory78,taylor92,kniffen97}.

Microquasars now join blazar AGNs as a firmly established 
class of very-high energy (VHE; with 
energies $\gtrsim$~a few hundred GeV -- TeV) $\gamma$-ray sources. 
The nonthermal continuum emission of blazars 
is believed to be produced in a relativistic plasma jet 
oriented at a small angle with respect to our line of sight. Their radio 
through UV/X-ray emission is most likely due to synchrotron emission by 
relativistic electrons in the jet, while the high energy emission can be 
produced by Compton upscattering of lower-energy photons off relativistic 
electrons \citep[for a recent review, see, e.g.][]{boettcher02}, or through 
hadronic processes \citep{mb92,ad01,muecke03}. 

Because of their apparent similarity with their supermassive AGN cousins, 
it has been suggested that Galactic microquasars may be promising sites 
of VHE $\gamma$-ray production \citep[e.g.,][]{bosch05a}. While earlier 
work on $\gamma$-ray emission from X-ray binaries focused on neutron star 
magnetospheres as $\gamma$-ray production sites 
\citep[e.g.][]{moskalenko93,moskalenko94,bednarek97,bednarek00},
the VHE $\gamma$-ray detection of LS~5039 suggests that the $\gamma$-ray
production is more likely to be associated with the jets. High-energy 
$\gamma$-rays of microquasars can be produced via hadronic 
\citep[e.g.][]{romero03} or leptonic processes. In the latter case,
the most plausible site would be close to
the base of the mildly 
relativistic jets, where ultrarelativistic electrons can Compton 
upscatter soft photons.
Possible sources of soft photons are the synchrotron radiation produced 
in the jet by the same ultrarelativistic electron population \citep[SSC =
synchrotron self-Compton;][]{aa99}, or external photon fields \citep{bp04,bosch05a}. 
Both LS~5039 and V615~Cas are high-mass X-ray binaries which are rather 
faint in X-rays, with characteristic 1 -- 10~keV luminosities of $\sim 
10^{34}$~ergs~s$^{-1}$. This is much lower than the characteristic bolometric 
luminosity of the high-mass companions of these objects, at $L_{\ast} \gtrsim 
10^{38}$~erg/s. Consequently, the dominant source of external photons in 
LS~5039 and V615~Cas is the companion's optical/UV photon field. 

The intense radiation field of the high-mass companion will, however, also
lead to $\gamma\gamma$ absorption of VHE $\gamma$-rays in the $\sim 100$~GeV
-- TeV photon energy range if VHE photons are produced close to the base
of the jet. For the case of LS~5039, \cite{aharonian05}
have estimated that VHE emission produced within $\sim 10^{12}$~cm of the
central engine of this microquasar would be heavily attenuated ($\tau_{\gamma\gamma}
\sim 20$ for $E_{\gamma} \sim 100$~GeV), and the observed spectrum of VHE 
photons would be hardened compared to its intrinsic shape. For comparison,
the $\gamma\gamma$ opacity due to stellar radiation back-scattered by the
stellar wind in the vicinity of the binary system (for which the angle of
incidence between the line of sight and the target soft photon field would
be more favorable for $\gamma\gamma$ absorption) may be estimated as
$\tau_{\gamma\gamma}^{\rm max} (r) \sim (L_{\ast} \, \dot M \, \sigma_T^2) / 
(2.7 \cdot 48 \, \pi^2 \, r^2 \, m_p \, v_{\infty} \, m_e c^3 \, \epsilon_{\ast})$,
where $L_{\ast}$ is the stellar luminosity, $\dot M$ is the mass outflow rate, 
$v_{\infty}$ is the terminal velocity of the stellar wind,
and $\epsilon_{\ast}$ is the dimensionless mean photon energy of the stellar
radiation field. For LS~5039, $L_{\ast} \sim 7 \times 10^{38}$~ergs~s$^{-1}$,
$m_e c^2 \, \epsilon_{\ast} \sim 3.5$~eV, $\dot M \approx 10^{-6.3} M_{\odot}$/yr,
and $v_{\infty} \approx 2500$~km~s$^{-1}$ \citep{mg02}, which yields
$\tau_{\gamma\gamma}^{\rm max} (r) \sim 0.1 / r_{12}^2$, where $r =
10^{12} \, r_{12}$~cm. Consequently, the $\gamma\gamma$ opacity will be
strongly dominated by the direct stellar photon field.
In the same spirit,
one can also estimate the effect of $\gamma$-ray absorption in the field of
atomic nuclei in the stellar wind. Using a total absorption cross section 
per unit mass of $\sigma_H \sim 0.012$~cm$^2$~g$^{-1}$ for GeV -- TeV 
$\gamma$-rays, we find $\tau_{\gamma Z} \approx 
\sigma_H \, \dot{M} 
/ (4 \pi \, v_{\infty} \, r_0) \approx 1.3 \times 10^{-4} 
/ r_{12}$ for LS~5093, which is also always much smaller than the 
$\gamma\gamma$ absorption depth.

In this {\it Letter}, we provide a more detailed analysis of the expected 
$\gamma\gamma$ absorption trough caused by direct companion star light, 
including its temporal modulation due to the orbital motion. The 
model description and the derivation of the general
expression for the $\gamma\gamma$ opacity as a function of $\gamma$-ray
photon energy and orbital phase is given in \S \ref{model}.
Numerical results for LS~5039 are presented in \S~\ref{results}.
\S~\ref{summary} contains a brief summary and discussion of our results.

\section{\label{model}Model description and analysis}

\begin{figure}[t]
\includegraphics[width=14cm]{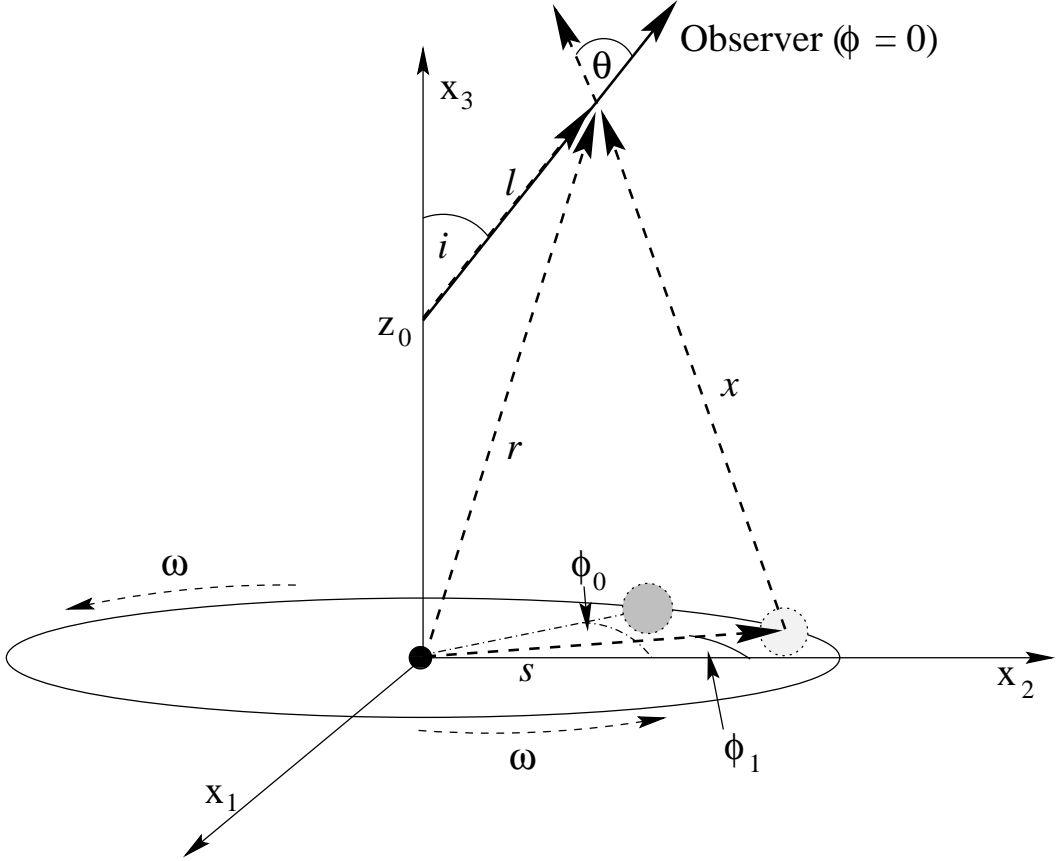}
\caption{Geometry of the model. The direction of the radio jets defines
the $x_3$ axis. The orbital plane of the binary system is the ($x_1$, $x_2$) plane, 
defined in such a way that line of sight towards the observer lies in the
($x_2$, $x_3$) plane, where the azimuthal angle $\phi = 0$.}
\label{geometry}
\end{figure}

We choose a generic model set-up as illustrated
in Fig.~\ref{geometry} \citep[for a comparable model setup in the AGN
case see][]{bd95}. The orbital plane of the binary system defines the
($x_1, x_2$) plane. The jet, assumed to be perpendicular to this plane, 
defines the
$x_3$ axis. The system is inclined with respect to our line 
of sight by
an inclination angle $i$. An azimuthal (phase) angle $\phi$ 
is defined such that $\phi = 0$ in the direction of the $x_2$ axis. The 
line of sight
lies in the $(x_2, x_3)$ plane. The $\gamma$-ray production 
site is
located at a height $z_0$ along the jet. The phase angle of the 
companion
star at the time of production of a $\gamma$-ray photon at $z_0$ 
is denoted $\phi_0$.
The distance the photon travels along the line of sight 
is denoted by $l$, whereas
a starlight photon travels a distance $x$ before 
interacting with the $\gamma$-ray
photon. The angle of incidence between 
the two photons is $\theta$, and $\mu \equiv \cos\theta$. The star was 
located at azimuthal angle $\phi_1$ at the
time when a photon, leaving 
the star at that time, interacts with a $\gamma$-ray
photon that has 
travelled a distance $l$ from $z_0$. This yields
$\phi_1 = \phi_0 + (2\pi / P)
\, (l - x) / c$, where $P$ is the orbital period.

With the definitions as shown in Fig.~\ref{geometry}, the distance x can
be calculated as $x = \vert {\bf x} \vert = \vert {\bf r} - {\bf s} \vert$,
where $s$ is the orbital separation of the binary system. We find
\begin{equation}
x^2 = s^2 + l^2 + z_0^2 + 2 \, l \, (z_0 \, \cos i - s \, \sin i \, \cos\phi_1).
\label{x2}
\end{equation}
From Fig.~\ref{geometry}, $\mu = {\bf l} \cdot {\bf x} / (l x)$, 
which yields

\begin{equation}
\mu = {l + z_0 \, \cos i - s \, \sin i \, \cos\phi_1 \over x}.
\label{mu}
\end{equation}

With these quantities, we can evaluate the $\gamma\gamma$ opacity of a
$\gamma$-ray photon with dimensionless energy $\epsilon_0 = E_0 / (m_e c^2)$
as

\begin{equation}
\tau_{\gamma\gamma} (\epsilon_0, z_0, \phi_0) = \int\limits_0^{\infty} dl \,
(1 - \mu) \int\limits_{2 \over \epsilon_0 \, (1 - \mu)}^{\infty} d\epsilon
\, \sigma_{\gamma\gamma} (\epsilon_0, \epsilon, \mu) \, n_{\rm ph}^{\ast}
(\epsilon, x)
\label{taugg}
\end{equation}
where
$\sigma_{\gamma\gamma}$ is the pair production cross section and

\begin{equation}
n_{\rm ph}^{\ast} (\epsilon, x) = {15 \over 4 \, \pi^5 \, m_e c^3} \,
{L_{\ast} \, \epsilon^2 \over \Theta_{\ast}^4 \, x^2 \, \left( e^{\epsilon
/ \Theta_{\ast}} - 1 \right)}.
\label{nph}
\end{equation}
The stellar spectrum
is approximated by a blackbody with dimensionless 
temperature $\Theta_* = kT_*/m_ec^2$,
and the star is approximated as a 
point source.

\section{\label{results}Results for LS~5039}

\cite{casares05} have analyzed optical intermediate-dispersion
spectroscopic observations of LS~5039, which provided an estimate
of the mass of the compact object of $M_X = 3.7^{+1.3}_{-1.0} \, 
M_{\odot}$, indicating
that it is likely to be a black hole. Other
relevant parameters of the binary system are $P = 3.91$~d, the 
luminosity of the
O6.5V type stellar companion, $L_{\ast} = 10^{5.3} 
\, L_{\odot}$, with an
effective surface temperature of $T_{\rm eff} 
= 39,000 \, ^o$K, a viewing angle of $i = 25^o$, and an orbital 
separation of $s \approx 2.5 \times 10^{12}$~cm \citep{casares05}. 
Note that the center of mass is rather close to the massive 
($M \approx 23 M_\odot$) stellar companion. However, because 
the light-travel and jet propagation times on the length scales investigated 
here ($\lesssim 10^{14}$~cm) are much shorter than the orbital period, the
detailed dynamics of the orbital motion are irrelevant for our purposes: 
at any given phase $\phi_0$, the geometry is basically stationary. An 
additional complication could be
introduced by the substantial eccentricity 
of the orbit, $e = 0.35$, with periastron at a phase angle of $\phi \sim 0.6 
\, \pi$
\citep{casares05}. However, in view of the still rather poor quality 
of the data currently available and expected in the near future, this may be 
regarded as a higher-order effect which may be introduced at a later stage.

\begin{figure}[t]
\includegraphics[height=14cm]{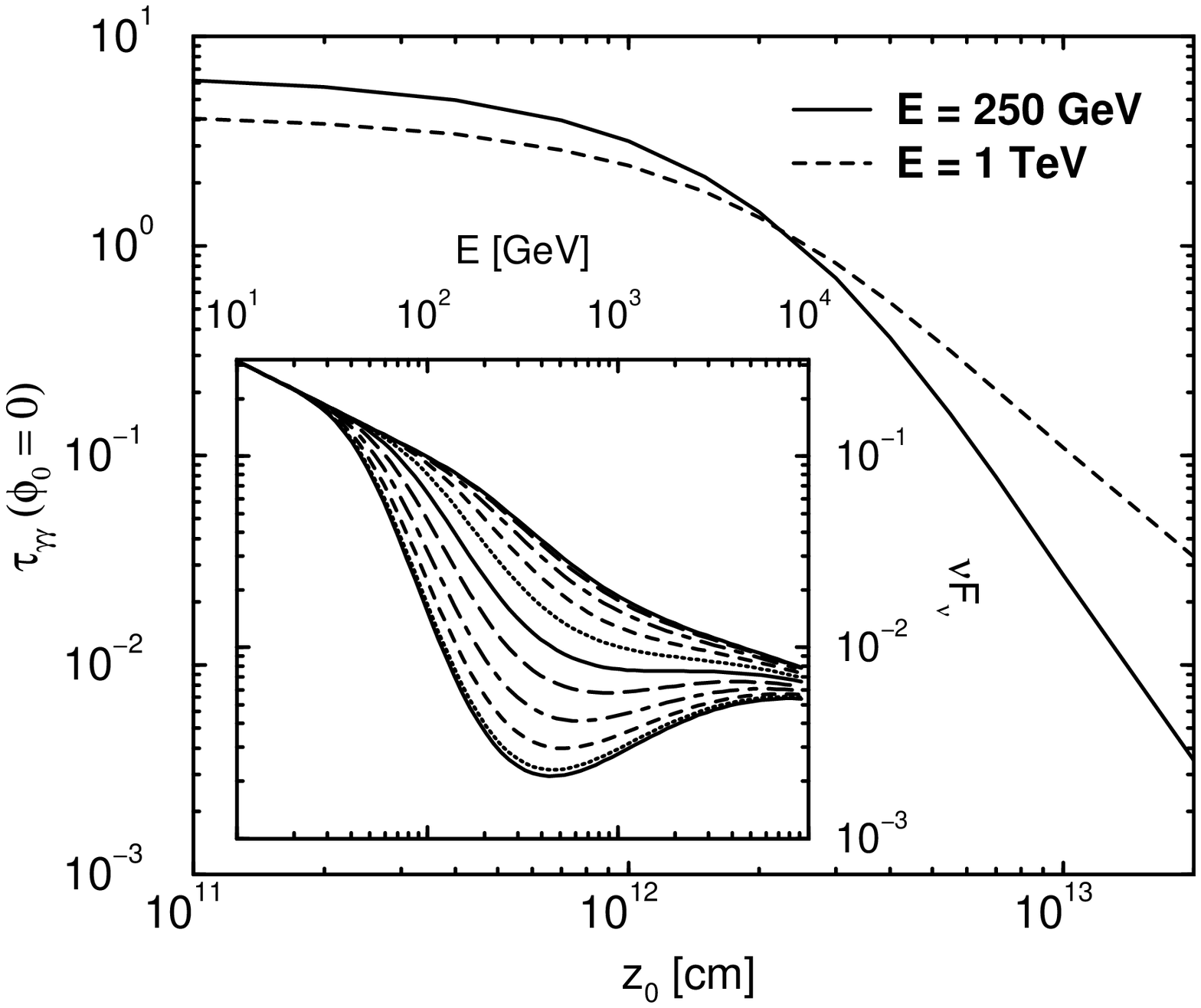}
\caption{$\gamma\gamma$ opacity at 250~GeV and 1~TeV as a function 
of the distance of the photon production region from the central 
compact object at phase $\phi_0 = 0$. The figure illustrates that 
(1) VHE photons produced
within a few $\times 10^{12}$~cm (i.e., 
of the order of the orbital
separation of the binary system) would 
be subject to substantial $\gamma\gamma$ absorption; (2) the minimum 
of the absorption trough
(maximum of $\tau_{\gamma\gamma}$ as a 
function of photon energy)
is shifting towards higher energies for 
larger distances from the
central source. 
Inset: Orbital modulation of the expected $\gamma\gamma$ absorption
trough, assuming a power-law spectrum with photon index
$\alpha_{\rm ph} = 2.5$ and a photon production site at $z_0 = 10^{12}$~cm.
The different curves represent the escaping photon spectrum at various
orbital phases, from $\phi_0 = 0$ (lowest curve) to $\phi_0 = \pi$ (highest
curve) in steps of $\pi/10$.}
\label{tauz}
\end{figure}

The inset to Fig.~\ref{tauz} illustrates the shape of the absorption trough
caused by $\gamma\gamma$ absorption and its dependence on the orbital phase.
Here we assume that the intrinsic $\gamma$-ray spectrum is a 
power-law with photon index $\alpha_{\rm ph} = 2.5$, and $z_0 = 10^{12}$~cm.
The various curves illustrate the orbital modulation of the absorption trough,
with the lowest (most heavily absorbed) curve corresponding to $\phi_0 = 0$
and the highest (least absorbed) curve corresponding to $\phi_0 = \pi$. The
modulation is a combined consequence of two effects: for phase angles closer 
to $\pi$, (a) the average distance
of the star to any point on the line of 
sight is longer and (b) the angle of
incidence $\theta$ is smaller. In addition
to effect (a) causing the overall photon number density of the stellar photon 
field to decrease, effect (b) causes the threshold for $\gamma\gamma$ pair 
production
to increase as $\epsilon_{\rm thr} = 2 / (\epsilon_{\ast} \, [1 - \mu])$. 
This leads to a decreasing overall depth of the absorption trough, and a shift 
of the minimum of the absorption trough towards higher photon energies. 

Fig.~\ref{tauz} shows the dependence of the absorption
feature on the location $z_0$ of the VHE $\gamma$-ray production site. The
$\gamma\gamma$ opacity is plotted for two photon energies, $E = 250$~GeV, and
$E = 1$~TeV at $\phi_0 = 0$. GeV -- TeV photons produced within $z_0 \sim s$ 
from the compact object
will be heavily attenuated for this phase angle. 
For photons produced at $z_0 \gg s$, $\gamma\gamma$ attenuation becomes
negligible. The crossing-over of the two curves in Fig.~\ref{tauz}
illustrates the shift of the minimum of the absorption trough to higher
photon energies with increasing distance from the central engine. This 
is a consequence of the decreasing incidence angle as discussed above. 

\begin{figure}[t]
\includegraphics[height=14cm]{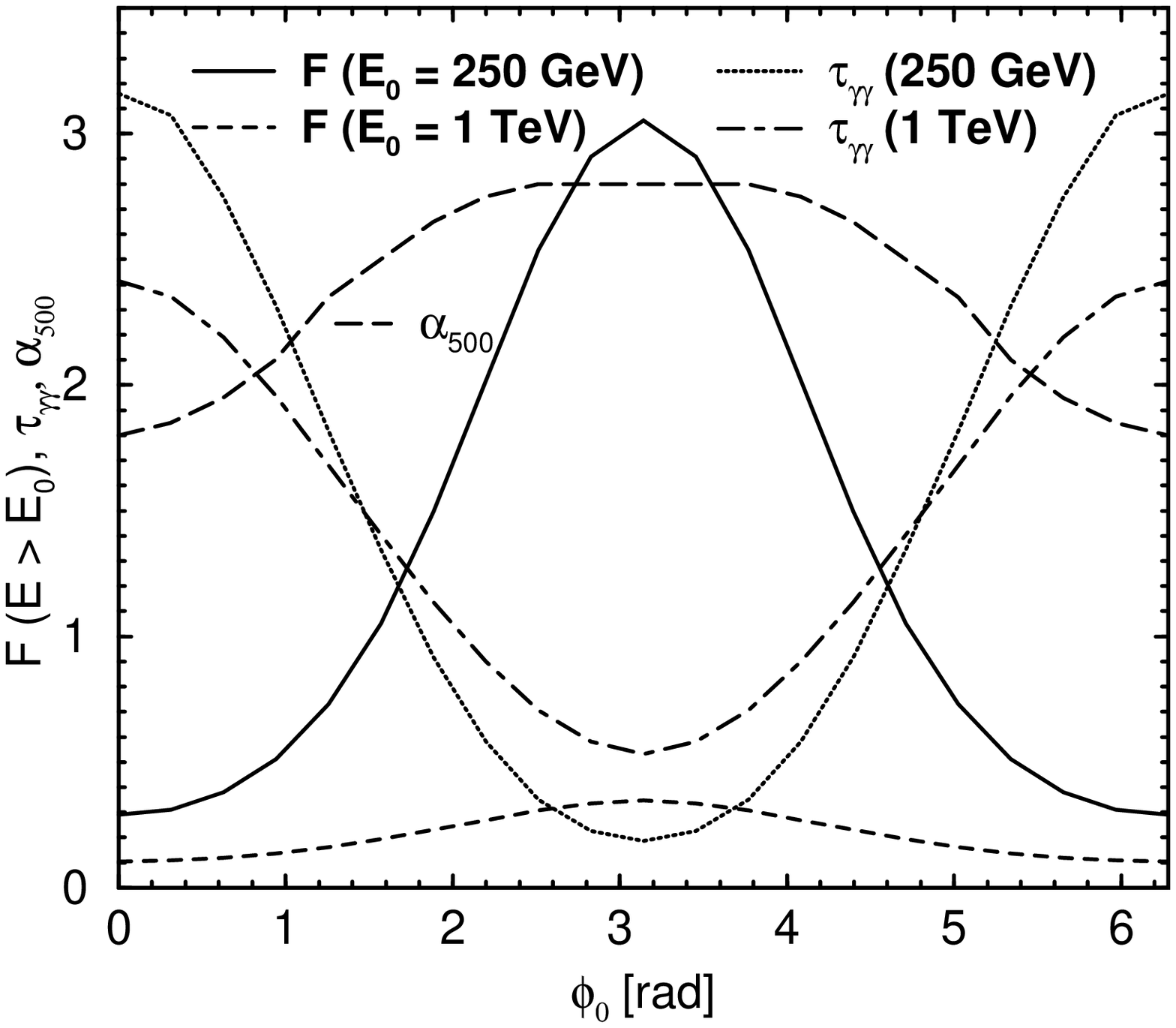}
\caption{Orbital modulation of the integrated photon number flux above 
energies $E_0 = 250$~GeV (solid) and $E_0 = 1$~TeV (short-dashed), 
the $\gamma\gamma$ opacity at $E = 250$~GeV (dotted) and $E = 1$~TeV
(dot-dashed), and the local photon spectral index $\alpha_{500}$ at 
500~GeV (long-dashed). As in the inset to Fig. \ref{tauz}, an underlying 
power-law of photon index $\alpha_{\rm ph} = 2.5$ and a photon production 
site at $z_0 = 10^{12}$~cm has been assumed. A periodic flux modulation is 
expected to be accompanied by positive spectral-index/flux correlation 
(spectral softening as the flux increases) at $E_0 \gtrsim 300$~GeV; the
opposite behavior is found at $E_0 \lesssim 100$~GeV.}
\label{flux_tau}
\end{figure}

Fig.~\ref{flux_tau} quantifies the orbital modulation of the $\gamma\gamma$
absorption trough for the same parameters as used in the inset to 
Fig.~\ref{tauz}. The dotted and
dot-dashed curves show the orbital modulation of the optical depth, illustrating
that (a) it is maximized for $\phi_0 \approx 0, 2\pi$, and 
(b) that the mimimum of the absorption trough shifts towards higher photon 
energies for phases closer to $\pi$. The most feasible way to detect an orbital
modulation would be to measure a periodicity of the observed photon flux. The
solid and short-dashed curves in Fig.~\ref{flux_tau} show the integrated 
photon number flux above energies $E_0 = 250$~GeV and
$E_0 = 1$~TeV, respectively. 
This illustrates that the relative modulation is
most pronounced at several 
hundred GeV, noting that the $\gamma\gamma$ absorption is largest 
at photon energies
$ E_1 \cong 2 \cdot 2 (m_ec^2)^2 / (k T_* [1 - 
\cos (\arctan(s/z_0) + i)])$. In this estimate, we have made use of the fact 
that the photon energy $\epsilon_{pk}$ at the peak of the $\gamma\gamma$ 
attenuation cross section is $\epsilon_{pk}\approx 2 \epsilon_{thr}$. For 
the case of LS 5039 with $z_0 = 10^{12}$ cm and $\phi_0 = 0$, $E_1 \cong 300$~GeV,
in good agreement with the inset in Fig.~\ref{tauz}. The spectral softening 
with
increasing flux is
also illustrated by the local photon spectral index $\alpha_{500}$ at 500~GeV, 
plotted as the long-dashed curve in Fig.~\ref{flux_tau}. The flux dip around
$\phi_0 \approx 0$ is accompanied by a hardening of the VHE $\gamma$-ray 
spectrum.

Because the bulk speed of microquasar jets is 
typically only mildly relativistic ($\Gamma \sim 2$), in dramatic contrast to 
blazars,
one might also need to 
consider the possibility of VHE $\gamma$-rays emanating from the counter jet. 
Thus, we have done a study as described above, for an inclination angle of
$i = 155^o$. While the results for this case are qualitatively very similar
to the $i = 25^o$ case, the resulting $\gamma\gamma$ opacities are typically
larger by a factor of $\sim 2$ -- 3 for photon production sites at $z_0 \lesssim
10^{12}$~cm, as expected because of the more favorable photon collision angle
and the line of sight passing by the star closer than for the approaching-jet
case. Because of the stronger absorption of $\gamma$ rays from the 
counter jet throughout the binary orbit,
and the reduction in the flux due to Doppler deboosting, the counter jet is
unlikely to enhance the $\gamma\gamma$ absorption signature
in the signal appreciably.
As for $i = 25^o$, $\gamma\gamma$ absorption effects become negligible for 
$z_0 \gg s$. 

\section{\label{summary}Summary and Discussion}

We have presented a detailed analysis of the effect of $\gamma\gamma$ absorption
of VHE $\gamma$-rays near the base of the jet of a microquasar by the intense
photon field of a high-mass stellar companion. We include the time-dependent,
periodic modulation of this effect due to the binary's orbital motion. We
applied our results to the specific case of LS~5039, which has recently been
identified as the counterpart of the VHE $\gamma$-ray source HESS~J1826-148.
Our results can be summarized as follows:

(1) VHE $\gamma$-rays produced closer to the central engine than
$z_0 \sim$~a~few~$\times 10^{12}$~cm, which is of the order of the
binary separation $s$, would be subject to very strong $\gamma\gamma$ 
absorption due to the stellar radiation field at orbital phases 
close to $\phi_0 = 0$.

(2) For VHE photon production sites at $z_0 \lesssim s$, the $\gamma\gamma$ 
opacity --- and, thus, the VHE $\gamma$-ray flux --- would be strongly
modulated on the orbital period of the binary system ($P = 3.91$~d in
the case of LS~5039). At orbital phases close to $\phi_0 = \pi$, the
intrinsic VHE $\gamma$-ray flux would still be virtually unabsorbed
even for $z_0 \sim 10^{12}$~cm. 

(3) The orbital modulation of the VHE $\gamma$-ray flux would be
characterized by a spectral hardening in the $\sim 300$~GeV -- 1~TeV
range during flux dips. At lower energies, the spectrum softens with
decreasing flux.

The HESS collaboration has reported no evidence for either flaring
or periodic variability from LS 5039 \citep{aharonian05}, 
although \citet{casares05} suggest that there is weak evidence 
for variability of the HESS emission with the orbital 
period. Periodic variability may also be indicated by X-ray 
observations \citep{bosch05b} of LS 5039. 

Besides $\gamma\gamma$ opacity effects, there are a few alternative
scenarios which might cause a periodic modulation of the $\gamma$-ray 
flux: 

(a) As a consequence of the substantial eccentricity of the orbit,
the rate of mass transfer from the stellar companion to the compact 
object, which is believed to be dominated by wind accretion, is likely 
to be periodically modulated. This modulation would also be expected to 
appear at radio and X-ray energies.

(b) Analogous to the phase-dependent modulation of the incidence angle 
for $\gamma\gamma$ absorption, this geometrical effect would also yield
a more favorable angle for Compton scattering of starlight photons into
the $\gamma$-ray regime at phases $\phi_0 \approx 0$.

(c) The orientation of the jet may also be mis-aligned with
respect to the normal of the orbital plane \citep{maccarone02,butt03}
and possibly precessing about the normal \citep{larwood98,torres05},
leading to additional modulations, including a changing Doppler boosting
factor.

Effect (a) would be expected to lead to an overall hardening
of the 
$\gamma$-ray spectrum at all energies with increasing $\gamma$-ray
flux, 
while (b) would lead to an overall softening throughout the GeV -- TeV
photon energy range because of the Klein-Nishina cutoff becoming noticeable
at lower observed photon energies with increasing flux. Both effects are
in contrast to the $\gamma\gamma$ absorption trough investigated in
this {\it Letter}. A stationary misalignment of the jet could lead to
a slight enhancement of the orbital modulation (if the jet makes a smaller
angle with the line of sight than the orbital-motion axis) or reduce it
(in the opposite case), but would not change our results qualitatively.
A $\gamma$-ray flux modulation due to jet precession can easily be 
disentangled from the orbital modulation since the precession period 
is generally different from the orbital period, so that its effect
would average out when folding observational data with the orbital period.
Consequently, a measurement of a non-thermal absorption trough at VHE 
$\gamma$-ray energies modulated with the orbital period would firmly 
establish the
importance of $\gamma\gamma$ absorption effects and thus 
place a robust
limit on the distance $z_0$ of the VHE $\gamma$-ray 
production site in LS 5039.

\acknowledgments
We thank Guillaume Dubus, Mathieu de Naurois, and Valenti Bosch-Ramon 
for helpful correspondence, and the referee for a helpful and constructive
report. This work was partially supported by NASA through XMM-Newton GO 
grant no. NNG~04GI50G, NASA INGEGRAL Theory grant no. NNG~05GK59G, and 
GLAST Science Investigation no. DPR-S-1563-Y. The work
of C.\ D.\ D.\ is 
supported by the Office of Naval Research.


\begin{thebibliography}

\bibitem[Aharonian et al.(2005)]{aharonian05}Aharonian, F., et al.,
2005, Science, 309, 746

\bibitem[Aharonian \& Atoyan(1999)]{aa99}Aharonian, F., \& Aharonian, A., 1999,
MNRAS, 302, 253

\bibitem[Atoyan \& Dermer(2001)]{ad01}Atoyan, A., \& Dermer, C.~D., 2001, 
Physical Review Letters, 87, 221102 

\bibitem[Bednarek(1997)]{bednarek97}Bednarek, W., 1997, A\&A, 322, 523

\bibitem[Bednarek(2000)]{bednarek00}Bednarek, W., 2000, A\&A, 362, 646

\bibitem[B\"ottcher(2002)]{boettcher02}B\"ottcher, 2002, in ``The 
Gamma-Ray Universe", proc. of the XXII Moriond Astrophysics Meeting, 
Eds. A. Goldwurm, D. N. Neumann, \& J. T. T. V$\hat{\rm a}$n, p. 151 

\bibitem[B\"ottcher \& Dermer(1995)]{bd95} B\"ottcher, M., \& Dermer, C.~D.\ 1995, \aap, 302, 37 

\bibitem[Butt(2003)]{butt03}Butt, Y., M., Maccarone, T. J., \& Prantzos, N.,
2003, ApJ, 587, 748

\bibitem[Bosch-Ramon \& Paredes(2004)]{bp04}Bosch-Ramon, V., \& Paredes,
J. M., 2004, A\&A, 417, 1075

\bibitem[Bosch-Ramon et al.(2005a)]{bosch05a}Bosch-Ramon, V., Romero, G. E., 
\& Paredes, J. M., 2005a, A\&A, 429, 267

\bibitem[Bosch-Ramon et al.(2005b)]{bosch05b} Bosch-Ramon, V., 
Paredes, J.~M., Rib{\' o}, M., Miller, J.~M., Reig, P., \& Mart{\'{\i}}, 
J.\ 2005b, \apj, 628, 388 

\bibitem[Casares et al.(2005)]{casares05}Casares, J., Rib\'o, M.,
Ribas, I., Paredes, J. M., Mart\'i, J., \& Herrero, A., 2005, MNRAS,
in press (astro-ph/0507549)

\bibitem[Gregory \& Taylor(1978)]{gregory78}Gregory, P. C., \& Taylor, A. R.,
1978, Nature, 272, 704

\bibitem[Kniffen et al.(1997)]{kniffen97} Kniffen, D.~A., et al.\ 1997, ApJ, 486, 126 

\bibitem[Larwood(1998)]{larwood98}Laarwood, J., 1998, MNRAS, 299, L32

\bibitem[Maccarone(2002)]{maccarone02}Maccaroone, T. J., 2002, MNRAS, 336, 1371

\bibitem[Mannheim \& Biermann(1992)]{mb92}Mannheim, K., \& Biermann, P. L.,
1992, A\&A, 253, L21

\bibitem[Mannheim(1993)]{mannheim93}Mannheim, K., 1993, A\&A, 221, 211

\bibitem[McSwain \& Gies(2002)]{mg02}McSwain, M. V., \& Gies, D. R., 2002, 
ApJ, 568, L27

\bibitem[Mirabel \& Rodr\'\i guez(1994)]{mirabel94}Mirabel, I. F., \&
Rodr\'\i guez, L. F., 1994, Nature, 371, 46

\bibitem[Moskalenko et al.(1993)]{moskalenko93}Moskalenko, I. V., Karakula, S.,
\& Tkaczyk, W., 1993, MNRAS, 260, 681

\bibitem[Moskalenko \& Karakula(1994)]{moskalenko94}Moskalenko, I. V., \&
Karakula, S., 1994, ApJS, 92, 567

\bibitem[M\"ucke et al.(2003)]{muecke03}M\"ucke, A., et al., 2003,
Astropart. Phys., 18, 593

\bibitem[Paredes et al.(2000)]{paredes00}Paredes, J. M., Mart\'\i, J., Rib\'o, M.,
\& Massi, M., 2000, Science, 288, 2340

\bibitem[Romero et al.(2003)]{romero03}Romero, G. E., Torres, D. F., Kaufman Bernad\'o,
M. M., \& Mirabel, I. F., 2003, A\&A, 410, L1

\bibitem[Taylor et al.(1992)]{taylor92}Taylor, A. R., Kenny, H. T. Spencer, R. E.,
\& Tzioumis, A., 1992, ApJ, 395, 268

\bibitem[Torres et al.(2005)]{torres05}Torres, D. F., Romero, G. E., Barcons, X.,
\& Lu, Y., 2005, ApJ, 626, 1015

17, 221

\end{thebibliography}
\end{document}